\documentclass[aps,prb,showpacs,two column,showkeys,superscriptaddress,preprintnumbers]{revtex4}
\usepackage{amsmath}
\usepackage{txfonts}
\usepackage[titletoc]{appendix}
\usepackage{amssymb}
\usepackage{array}
\usepackage{mathrsfs}
\usepackage{subfigure,overpic}
\usepackage{dcolumn}

\newcommand{\PreserveBackslash}[1]{\let\temp=\\#1\let\\=\temp}
\newcolumntype{C}[1]{>{\PreserveBackslash\centering}p{#1}}
\newcolumntype{R}[1]{>{\PreserveBackslash\raggedleft}p{#1}}
\newcolumntype{L}[1]{>{\PreserveBackslash\raggedright}p{#1}}

\begin{document}

\title{Signature of topological phase transition in the RKKY interaction of
silicene}
\author{Hou-Jian Duan, Shi-Han Zheng, Zhen-Long Sun, Mou Yang, and Rui-Qiang
Wang}
\email{rqwanggz@163.com}
\affiliation{Laboratory of Quantum Engineering and Quantum Materials, \\
School of Physics and Telecommunication Engineering, South China Normal
University, Guangzhou 510006, China}

\begin{abstract}
Silicene offers an ideal platform for exploring the phase transition due to
strong spin-orbit interaction and its unique structure with strong
tunability. With applied electric field and circularly polarized light,
siliccene is predicted to exhibit rich phases. We propose that these
intricate phase transitions can be detected by measuring the bulk
Ruderman-Kittel-Kasuya-Yosida (RKKY) interaction. We have in detail analyzed
the dependence of RKKY interaction on phase parameters for different
impurity configurations along zigzag direction. Importantly, we present an
interesting comparison between different terms of RKKY interaction with
phase diagram. It is found that the in-plane and out-of-plane terms can
exhibit the local extreme value or change of sign at the phase critical
point and remarkable difference in magnitude for different phase regions.
Consequently, the magnetic measurement provides unambiguous signatures to
identify various types of phase transition simultaneously, which can be
carried out with present technique.
\end{abstract}

\maketitle


Topological quantum phase transition has received great interest in
condensed matter of states for searching for new matter states\cite{p93},
such as very recently emerging topological insulators (TIs), Weyl or Dirac
semimetals. Topological quantum state possesses many exotic and robust
properties with potential application in quantum calculations\cite{has}.
Topological phases are usually classified with topological indices. In 2D
quantum system, the topological indices are reduced to the charge- and
spin-Chern numbers\cite{has,spin chern number}, obtained by summation over
the Berry curvature. Nevertheless, how to identify these different
topological states experimentally is a challenging problem. The most
instinctive method to detect a topological phase is to measure the
spin-resolved quantum Hall conductivity or to directly probe topological
states. However, these electric measurements are difficult to perform in
quantum Hall systems and moreover topological edge states are easy to suffer
from the disturbance from bulk states which are unavoidable due to the
existence of imperfections in the composition.

Much effort is made to find other new tools for probing the topological
phase transition. The phase-dependent heat currents provide a robust tool to
distinguish the existence of topological Andreev bound states from trivial
Andreev bound states in superconductor/TI Josephson junction\cite{p94}. To
explore the existence of fractional quantum Hall states in TIs, the authors%
\cite{p245} presented thermoelectric measurements on the $\mathrm{Bi_2Te_3}$
crystal. The magnetic susceptibility of electrons was studied in topological
nodal semimetals, in which a giant anomaly is regarded to be useful in
experimental identification of the Weyl, Dirac and line node semimetals\cite%
{p195}. The spin response in HgTe quantum wells\cite{p021} reveals that
unconventional spin-related properties can distinguish the paradigmatic TI
material from the other 2D electronic systems.

Silicene, a single layer of silicon atoms with a planar honeycomb lattice
structure\cite{vot}, offers an ideal platform for exploring the phase
transition. Besides large spin-orbit interaction up to $3.9$ meV\cite{liu1},
silicene possesses a buckled hexagonal structure, in which two atoms in the
translational unit cell reside on different planes, making its bandgap
tunable easily by applying an electric field perpendicular to the silicene
sheet\cite{dru1}. The electric field breaks inversion symmetry while the
circularly polarized light breaks time-reversal symmetry, both of which
modify the Berry curvatures in the momentum space so that the occupied
electronic states change the topological properties\cite{p026}. When both of
fields are applied, the silicene is predicted to exhibit rich phases:
quantum spin Hall insulator (QSHI), conventional bulk insulator (CBI),
photoinduced quantum Hall insulator (P-QHI), and photoinduced spin-polarized
quantum Hall insulator (PS-QHI)\cite{p026,spin chern number}. It is an
intriguing problem how to detect experimentally which phase the system stays
in just by the bulk property. For this, Ezawa\cite{zawa,zawa2,zawa3} has
proposed methods to differentiate the QSHI from the CBI phase by measuring
the diamagnetism or circular dichroism. To probe more intricate phase
transitions, Jin $et$ $al.$\cite{jin} have suggested to measure the Nernst
conductivity, from which phase boundaries can be determined by comparison
the charge- with spin-Nernst conductivities.

In this Letter, we propose that these intricate phase transitions in
silicene can be detected by measuring the bulk Ruderman-Kittel-Kasuya-Yosida
(RKKY). The RKKY interaction, which describes the indirect exchange coupling
between magnetic impurities mediated by the itinerant electrons, greatly
depends on the spin-orbit interaction of host materials\cite{jpcm,p045,ima}. Meanwhile, the
spin-orbit interaction plays a vital role in topological phase transitions.
Thus, it is natural to expect that there is a close relation between the
RKKY interaction and phase transition. We have in detail
analyzed dependence on phase parameters of RKKY interaction and present a
RKKY phase diagram. It is shown that magnetic measurement, even for the bulk
states, will provide information enough to determine various phase
boundaries and identify different phases.

\emph{Model and Method}--Silicene has a honeycomb lattice with two different
atoms in the translational unit cell. Employing the tight-binding model for
the four bands\cite{zawa,zawa2}, the Hamiltonian is given by
\begin{equation}
H=-t\sum\limits_{\langle i,j\rangle s}c_{is}^{+}c_{js}+i\frac{\lambda _{so}}{%
3\sqrt{3}}\sum\limits_{\left\langle \left\langle i,j\right\rangle
\right\rangle ss^{\prime}}c_{is}^{+}\mathbf{\sigma }_{ss^{\prime }}\cdot (%
\mathbf{d}_{i}\times \mathbf{d}_{j})c_{js^{\prime }}+U\sum\limits_{is}\mu
_{i}c_{is}^{+}c_{is}  \label{Hamiltonian}
\end{equation}%
where $\left\langle i,j\right\rangle $ ($\left\langle \left\langle
i,j\right\rangle \right\rangle $) runs over the nearest-neighbor
(next-nearest-neighbor) hopping sites, $c_{is}^{+}$ creates an electron with
spin $s$ at site $i$, $\mathbf{\sigma }$ is the Pauli matrix of spin, $%
\mathbf{d}_{i}$ and $\mathbf{d}_{j}$ are the in-plane unit vectors along
which the electron traverses from site $j$ to $i$. The first two terms
describe the silicene with hopping energy $t=1.6$ $\mathrm{eV}$ and the
intrinsic spin-orbit coupling $\lambda _{so}\approx 3.9\mathrm{meV}$\cite%
{liu1,spin orbital1,spin orbital2}, while the weak Rashba spin-orbital
interaction is neglected\cite{p026}. The third term stands for the staggered
potential with $\mu _{i}=\pm 1$ for $A$ ($B)$ site and $U=E_{z}d/2$, caused
by an electric filed $E_{z}$ exerting on the buckled lattice structure\cite%
{tak}, where two sublattice planes are separated by a distance of $d=0.46%
\mathring{A}$. By transforming Eq. (1) into the momentum space and then
expanding it at the two Dirac points $\mathbf{K}_{\eta }$ ($\eta =\pm $) in
the Brillouin zone (BZ), we in the pseudospin space $\{A,B\}$ obtain the
low-energy Dirac Hamiltonian
\begin{equation}
H_{\eta s}=\left(
\begin{array}{cc}
U_{\eta s} & \hbar v_{F}k\Phi _{K_{\eta }} \\
\hbar v_{F}k\Phi _{K_{\eta }}^{\ast } & -U_{\eta s}%
\end{array}%
\right) .
\end{equation}%
Here, $v_{F}=\frac{\sqrt{3}}{2}at,$ $U_{\eta s}=-s\lambda _{so}\eta -U$ with
$s,\eta =\pm 1$ are the spin and valley indices, respectively, and $\Phi
_{K_{\eta }}=\eta e^{-i\pi /3+i\eta \theta }$ with the polar angle $\theta
=\arctan \left( k_{y}/k_{x}\right) $ and an extra phase factor\cite{rkky
graphene3} stemming from the specific $K_{\eta }$.

In order to present rich phases, we assume the silicene sheet is in addition
irradiated by a beam of circularly polarized light. The photoinduced effect
is considered by the Peierls substitution $\hbar \mathbf{k}$$\rightarrow$$%
\hbar \mathbf{k}+$ $e\mathbf{A}\left( t\right)$, where vector potential $%
\mathbf{A}\left( t\right) =A\left( \sin \omega t,\cos \omega t\right) $ is a
periodic function of time $T=2\pi /\omega$ with $\omega $ being the light
frequency. By using the Floquet theory\cite%
{p026,floquet1,floquet2,floquet3,floquet4,floquet5}, the time dependence can
be mapped to a Hilbert space of time-independent multi-photon Hamiltonian.
For the off-resonant light with the high-frequency limit $A^{2}/\omega \ll 1$%
, one can decouple the zero-photon state from the other states and only
consider its dressed effect through second-order virtual photon absorption
and emission processes\cite{p026,floquet3,ope,li}. As a consequence, the
modified part of Hamiltonian by light reads $V_{n}=\left[ V_{-1},V_{+1}%
\right] /\hbar \omega +O(A^{4})$ with $V_{n}=\frac{1}{T}%
\int_{0}^{T}H(t)e^{-in\hbar \omega t}dt$ and the effective Hamiltonian is
approximately expressed as
\begin{equation}
H_{\eta s}^{\prime }=H_{\eta s}+V_{n=0}=H_{\eta s}+\Omega \sigma _{z},
\end{equation}%
with the illumination parameter $\Omega =\frac{3t^{2}A^{2}}{4\hbar \omega }.$
By diagonalizing the Hamiltonian $H_{\eta s}^{\prime }$, the low-energy
dispersion reads
\begin{equation}
E_{\eta s}^{\pm }=\pm \sqrt{\hbar ^{2}v_{F}^{2}k^{2}+U_{\eta s}^{2}}
\end{equation}%
where the energy gap $2\left\vert U_{\eta s}\right\vert =2\left\vert \left(
\Omega -s\lambda _{so}\right) \eta -U\right\vert $ can\ be opened or closed,
controlled by both the light and electric fields. Consequently, the
topological phase transition occurs among four categories\cite{p026}: P-QHI,
QSHI, PS-QHI, and CBI.

We assume two magnetic impurities $\mathbf{S}_{i}$ placed on the lattice
sheet interacting with conducting electrons via $H_{int}=\lambda \sum_{i}%
\mathbf{S}\left( \mathbf{r}_{i}\right) \cdot \mathbf{s}\left( \mathbf{r}%
_{i}\right) $, where $\mathbf{S}(\mathbf{r}_{i})$ [$\mathbf{s}(\mathbf{r}%
_{i})$] is the spin of impurities (itinerant electrons) and $\lambda $ is
the spin-exchange coupling strength. For weak coupling, we can replace $%
H_{int}$ with the RKKY interaction, which in the second-order perturbation
theory\cite{ima,perturbation theory1,perturbation theory2,perturbation
theory3,perturbation theory4} is given by
\begin{equation}
H_{RKKY}^{\alpha \beta }=\frac{-\lambda ^{2}}{\pi }\mathrm{Im}\int_{-\infty
}^{E_{F}}\mathrm{Tr}\left[ \left( \mathbf{S}_{1}\cdot \sigma \right)
G_{\alpha \beta }\left( \mathbf{R},\varepsilon \right) \left( \mathbf{S}%
_{2}\cdot \sigma \right) G_{\beta\alpha }(-\mathbf{R},\varepsilon )\right]
d\varepsilon .
\end{equation}%
Here, $\alpha ,\beta =\{A,B\},$ $\mathbf{R}$ is spatial distance between two
impurities, $E_{F}$ is Fermi level, and the trace is over the spin degree of
freedom. The retarded Green's function $G_{\alpha \beta }(\mathbf{R}%
,\varepsilon )=\sum_{\eta }\int e^{-i\mathbf{k\cdot R}}d^{2}\mathbf{k}\left[%
1/(\varepsilon +i0^{+}-H_{\eta s}^{\prime }\right]_{\alpha \beta }$ is a $%
2\times 2$ matrix in spin space. In next discussions, we focus on the
impurities placed on the same sublattice (e.g., $\alpha=\beta=A$) and drop
the subscript for briefness. Consequently, the matrix element of Green's
function is diagonal in spin space and reads
\begin{equation}
G^{s,s^{\prime}}\left( \mathbf{R},\varepsilon \right) =-\frac{%
2\pi\sigma_{s,s^{\prime}} }{\varsigma \hbar ^{2}v_{F}^{2}}\sum\limits_{\eta
=\pm 1}e^{i\mathbf{K}_{\eta }\mathbf{R}}\left( \varepsilon +U_{\eta
s}\right) K_{0}\left( \mathscr{R}_{U_{\eta s}}\right) ,  \label{GAA}
\end{equation}%
where $K_{0}\left( x\right) $ is the modified Bessel function of the second
kind, $\varsigma$ is the area of BZ, and $\mathscr{R}_{x}{=R\sqrt{%
x^{2}-\varepsilon ^{2}}/\hbar v_{F}}$ with $R=|\mathbf{R}|$. By inserting
the above Green's functions in Eq. (5), the RKKY interaction can be
rewritten as
\begin{equation}
H_{RKKY}=J_{\Vert
}\sum\limits_{i=x,y}S_{1i}S_{2i}+J_{z}S_{1z}S_{2z}+J_{DM}\left( \mathbf{S}%
_{1}\times \mathbf{S}_{2}\right) _{z},
\end{equation}%
which is divided into three terms according to the polarizations of the
impurities.

\emph{RKKY under light field}--To detect the topological phases, we expect
to search for signatures of the RKKY interaction characterizing the
phase-transition point and various phase regions. Firstly, we consider the
case of silicene sheet irradiated by a beam of off-resonant light but in the
absence of electric field. The light field breaks the time-reversal symmetry
and so causes spin splitting $\left\vert \Omega \pm \lambda _{so}\right\vert
$ in the energy spectrum from the original spin-degenerate bands $s=\pm 1$.
With the increase of light strength, the bandgap is closed first at the
critical point $\Omega =\pm \lambda _{so}$ and then enters a new topological
phase of P-QHI from QSHI state. Different topological phases can be
clarified by topological quantum numbers ($C,C_{s}$), corresponding to
charge- and spin-Chern numbers, respectively. They are usually defined as $C=
$ $C_{\uparrow }+C_{\downarrow }$ and $C_{s}=(C_{\uparrow }-C_{\downarrow
})/2$ and calculated with the integral of a closed path $C_{s}=\frac{1}{2\pi
}\sum\limits_{n}\int_{BZ}d\mathbf{k}\Omega _{xy}^{n}(\mathbf{k})$ over the
Berry curvature $\Omega ^{n}(\mathbf{k})$ of the $n$-th band\cite{117}. In
Fig. 1, two phase regimes of the QSHI ($0,1$) and P-QHI\ ($-2,0$) are
divided by a vertical dotted line. In only irradiation of light, the bandgap
is reduced to $\left\vert V_{s}\left( \Omega \right) \right\vert $, where
the short-hand notation is for $V_{s}\left( x\right) =x+s\lambda _{so}$, and
the various terms of the RKKY is derived as $J_{i}=-2C\int_{-\infty
}^{E_{F}}N_{i}d\varepsilon $ $\left( C=8\pi \lambda ^{2}/\varsigma ^{2}\hbar
^{4}v_{F}^{4}\right) $ with
\begin{eqnarray}
N_{\Vert } &=&2\left[ \varepsilon ^{2}\cos ^{2}\left( \frac{\mathbf{1}}{2}%
\Delta \mathbf{K}\cdot \mathbf{R}\right) +\sin ^{2}\left( \frac{\mathbf{1}}{2%
}\Delta \mathbf{K}\cdot \mathbf{R}\right) \prod\limits_{s=\pm }V_{s}\left(
\Omega \right) \right]   \notag \\
&&\times \prod\limits_{s=\pm }K_{0}\left[ \mathscr{R}_{V_{s}\left( \Omega
\right) }\right] , \\
N_{z} &=&\sum\limits_{s=\pm }\left[ \varepsilon ^{2}\cos ^{2}\left( \frac{%
\mathbf{1}}{2}\Delta \mathbf{K}\cdot \mathbf{R}\right) +\sin ^{2}\left(
\frac{\mathbf{1}}{2}\Delta \mathbf{K}\cdot \mathbf{R}\right) V_{s}^{2}\left(
\Omega \right) \right]   \notag \\
&&\times K_{0}^{2}\left[ \mathscr{R}_{V_{s}\left( \Omega \right) }\right] ,
\\
N_{DM} &=&-2\lambda _{so}\varepsilon \sin (\Delta \mathbf{K}\cdot \mathbf{R}%
)\prod\limits_{s=\pm }K_{0}\left[ \mathscr{R}_{V_{s}\left( \Omega \right) }%
\right] ,
\end{eqnarray}%
where $\Delta \mathbf{K=K-K^{\prime }}$ is deference of momentum for any two
adjacent Dirac points in BZ. We choose two valleys at $\mathbf{K(\mathbf{K}%
^{\prime })=}\frac{2\pi }{3a}(\pm 1,\sqrt{3}).$ Obviously, due to the
oscillation factor $\cos (\Delta \mathbf{K}\cdot \mathbf{R})$ or sin$(\Delta
\mathbf{K}\cdot \mathbf{R}),$ the RKKY interaction is closely related to
spatial distance $\mathbf{R}$ between impurities. While the impurity
distance fulfils $\mathbf{R}=na\hat{x}$ along the zigzag direction, the
oscillating part $\sin (K_{x}R_{x})$ repeats three values: $\frac{\sqrt{3}}{2%
}$, $-\frac{\sqrt{3}}{2}$, and $0$, corresponding respectively to the
impurity configuration satisfied $\mathrm{Mod}(R/a,3)=1,2,0$. This is
indicated by $A_{1}$, $A_{2}$ and $A_{3}$ in inset of Fig. 1 while the other
impurity is fixed at $A_{0}$ point. However, $\sin (\Delta \mathbf{K}\cdot
\mathbf{R)}$ always vanishes in the armchair direction, making the RKKY
featureless, so we in the following focus on the impurities distributed
along the zigzag direction and the system is half filled $(E_{F}=0)$.

\begin{figure}[tbp]
\centering \includegraphics[width=0.49\textwidth]{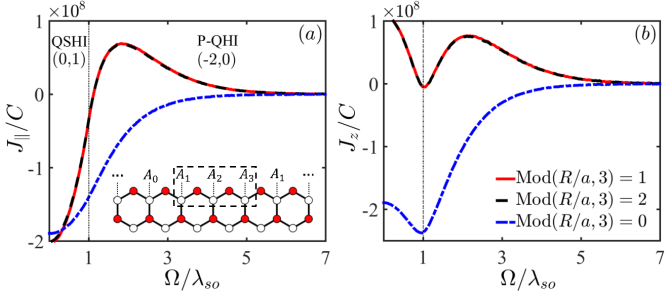}
\caption{(Color online) The variation of RKKY exchange coupling with
illumination parameter $\Omega$. The QSHI and P-QHI phases are divided by a
vertical dotted line. Two impurities are distributed on the same lattice
along the zigzag direction, as shown in inset, with three configurations in
spatial distance $R=270a$ [$\mathrm{Mod}(R/a,3)=0$], $271a$ [$\mathrm{Mod}%
(R/a,3)=1$] and $272a$ [$\mathrm{Mod}(R/a,3)=2$].}
\end{figure}
We in Fig. 1 present the numerical results for the illumination dependence
of different terms of the RKKY interaction in the long range for three types
of impurity positions. For the distances satisfying $\mathrm{Mod}(R/a,3)=1,2$%
, there emerges a prominent signature in Fig. 1 (a) that the in-plane term $%
J_{\Vert }<0$ is ferromagnetic in the QSHI phase while it changes to be
antiferromagnetic in the P-QHI phase. Interestingly, the transition point is
close to the critical value of phase $\Omega =\lambda _{so}$. This behavior
can be understood from Eq. (8), where the second term in $N_{\Vert }$ plays
a dominant role near the critical point and the sign of its integral is
almost determined by $V_{+}(\Omega )V_{-}(\Omega )=\Omega ^{2}-\lambda
_{so}^{2}$, namely, for QSHI with $|\Omega |<\lambda _{so}$ the value of $%
J_{\Vert }$ is negative while it is positive otherwise. For the impurity
configuration of $\mathrm{Mod}$$(R/a,3)=0$, no such sign is observable due
to $\sin (\Delta \mathbf{K}\cdot \mathbf{R/2})=0$. Besides, it is very
interesting to find that the out-plane term $J_{z}$ in Fig. 1(b) provides
more accurate signature of phase transition, manifesting itself by a large
dip exactly at the critical point. This dip structure occurs for all of
three impurity configurations, independent of the distance of impurity as
long as in the long range. After replacing the Bessel function $K_{0}(x)$
with $\sqrt{\pi /2x}e^{-x}$ in the long range\cite{Saremi} under
consideration and taking a derivative of the $N_{z}$ with respect to $\Omega
$, we obtain a result in the form of $dN_{z}/d\Omega \propto (\Omega
-\lambda _{so})f(\Omega ,\varepsilon )$, which explains the dip feature.
Although $J_{z}$ cannot changes sign like $J_{\Vert }$ when the phase
transition happens, its magnitude is quantitatively different in QSHI and
P-QHI phases. For the DM term $J_{DM}$, it keeps vanished for the Fermi
energy $E _{F}=0$ due to the electron-hole symmetry and the well-preserved
inversion symmetry\cite{Chang}.

\emph{RKKY under electric field}--We here discuss the variation of the RKKY
interaction when the silicene is subject to a perpendicular electric field $U
$. As $|U|>\lambda _{so}$, the resulting staggered potential can drive the
silicene from QSHI phase to CBI phase, whose topological numbers are
labeled, respectively, as $(0,1)$ and $(0,0)$ in Fig. 2. This topological
phase transition is discussed in detail in Ref.\cite{p026, dru1}. For this
case, we derivate the RKKY interaction as $J_{i}=-C\int_{-\infty
}^{E_{F}}N_{i}d\varepsilon $ with
\begin{eqnarray}
N_{\Vert } &=&2\prod\limits_{s=\pm }\zeta _{s}+\cos (\Delta \mathbf{K}\cdot
\mathbf{R})\sum\limits_{s=\pm }\zeta _{s}^{2}, \\
N_{z} &=&\sum\limits_{s=\pm }\zeta _{s}^{2}+2\cos (\Delta \mathbf{K}\cdot
\mathbf{R})\prod\limits_{s=\pm }\zeta _{s}, \\
N_{DM} &=&\sin (\Delta \mathbf{K}\cdot \mathbf{R})\sum\limits_{s}s\zeta
_{s}^{2},
\end{eqnarray}%
where $\zeta _{s}=\left[\varepsilon-V_{s}\left( U\right)\right] K_{0}\left( %
\mathscr{R}_{V_{s}\left( U\right) }\right) $.

\begin{figure}[tbp]
\centering \includegraphics[width=0.49\textwidth]{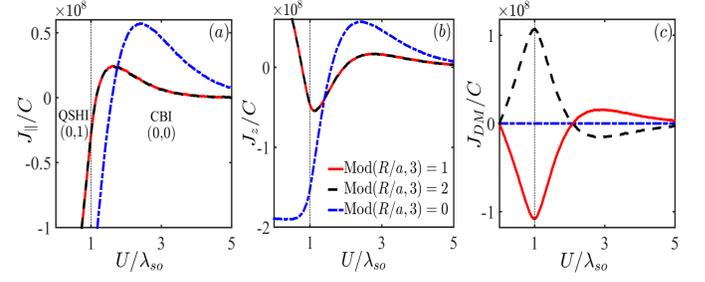}
\caption{(Color online) The dependence of (a) $J_{\parallel}$, (b) $J_{z}$,
and (c) $J_{DM}$ on the electric potential $U$. The others are the same as
in Fig. 1. }
\end{figure}
Performing the numerical calculations with above expressions, we plot the $%
J_{\Vert },J_{z}$ and $J_{DM}$ terms of the exchange coupling in Figs.
2(a)-(c), respectively. For two impurities placed at $\mathrm{Mod}$$(R/a,3)=0
$, though $J_{\Vert }$ and $J_{z}$ present a transition from the
ferromagnetic to antiferromagntic phase, the transition point is far away
from the critical point $U=\lambda _{so}$. In contrast, both $J_{\Vert}$ and
$J_{z}$ for impurity configuration $\mathrm{Mod}$$(R/a,3)=1,2$ provide a
relatively accurate signature for phase boundary: a
ferro-to-antiferromagntic transition for $J_{\Vert }$ and a dip structure
for $J_{z}$. They are approximately located at the phase transition point.
Very different from the case of light irradiation, $J_{DM}$ shows a strong
dependence on the electric field as in Fig. 2(c), where $\mathrm{Mod}%
(R/a,3)=1,2$ exhibit a dip and a peak, respectively, providing an
unambiguous fingerprint to ascertain the phase boundary between QSHI and CBI.

\begin{figure}[tbp]
\centering \includegraphics[width=0.49\textwidth]{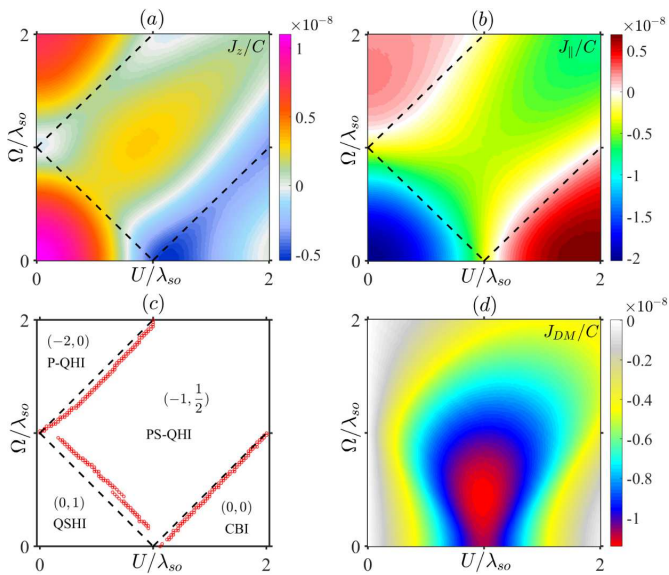}
\caption{(Color online) The phase diagrams of (a) $J_{z}$, (b) $J_{\parallel}
$, and (d) $J_{DM}$ as functions of $U$ and $\Omega$. (c) The comparison
between the boundary (black dashed lines) of different topological phase
transitions and signature of $J_{z}$ (red circles) which is selected from
the local minimum value in (a).}
\end{figure}
\emph{RKKY under both electric and light fields}--When both the electric and
light fields are exerted, there emerge rich phases: QSHI, P-QHI, PS-QHI, and
CBI as shown in Fig. 3(c), where the dashed lines denote the phase
boundaries. Since the expressions are too tedious, we here only give the
numerical results of $J_{z},J_{\Vert },$ and $J_{DM}$ for $\mathrm{Mod}$$%
(R/a,3)=1$ as functions of the electric potential $U$ and illumination
parameter $\Omega $ in Figs. 3(a), (b) and (d), respectively. Intriguingly,
the phase plots in Figs. 3(a) and (b) present distinct changes in color in
different regions, which can be used to differentiate the different phases
though it is not too very strict. Importantly, $J_{z}$ not only has
different values for different states, but also clearly characterizes the
various phase boundaries, especially for the phase transitions between
PS-QHI and CBI, PS-QHI and P-QHI, and QSHI and P-QHI, where a largest dip
exists. To compare with the phase plot, we describe the characterizing
signatures of the RKKY interaction in Fig. 3(c), marked with red circles by
selecting the local minimal values in their boundaries. With a tolerable
error, dependence of $J_{z}$ on electric and light fields provides
unambiguous signatures to identify the various phase transitions. By
comparison, the phase boundaries of $J_{\Vert }$ in Fig. 3(b) become blurry
but show remarkable difference in magnitude or sign for different phase
regions, suitable for characterizing different phase regions. It is noted
that, $J_{DM}$ in Fig. 3(d) with a deep dip exactly at the critical point
can only be applied to divide the phase transition between QSHI and CBI
states, but cannot characterize the other intricate phases. As discussed
above, the main reason is that $J_{DM}$ is insensitive to irradiation.
Therefore, the measurement of $J_{z}$ as well as $J_{\Vert }$ could be a
valid method to divide the different topological areas and their phase
boundaries.

\emph{Summary}--We have studied the RKKY coupling of a monolayer silicene
subject to an off-resonant light and a perpendicular electric field. Due to
topological phase transition, the RKKY coupling shows strong dependence on
the illumination and electric potential. Based on the lattice Green's
function formalism\cite{hor}, we have analyzed in detail the variation of
the RKKY interaction for different impurity configurations along zigzag
direction. It is found that the indirect magnetic interaction has tight
connection with various topological phase transitions. For the case
irradiated by light, a dip structure of $J_{z}$ can exactly identify the
phase transition of QSHI/P-QHI while the peak or dip of $J_{DM}$ can feature
the critical point of phase transition of QSHI/CBI induced by an electric
field. For more complex phase driven by both light and electric fields, it
is found that $J_{z}$ provides information enough to divide the different
topological areas with a forgivable error in the phase boundary. Also, $%
J_{\Vert }$ exhibits remarkable difference of magnitude or sign in different
phase regions though it is hard to differentiate the phase boundary. Since
there are quite rare methods to detect them, especially for the phase
transition between PS-QHI and P-QHI, measurement on the RKKY interaction
provides us an alternative method to probe the rich topological phases in
silicene or other spin-orbit systems. The underlying physics is that both
the topological property and magnetic property are determined by bandgap of
the band structure. Our proposal is expected to feasible with present
technique of spin-polarized scanning tunneling spectroscopy\cite{zhou},
which can measure the magnetization curves of individual atoms.

\acknowledgements This work was supported by NSF of China Grant Nos.
11474106 and 11274124, as well as by the Innovation Project of Graduate
School of South China Normal University.

\end{document}